\documentclass[amsmath,amssymb,superscriptaddress,nobalancelastpage,aps,longbibliography,twocolumn,showpacs]{revtex4-1}

\usepackage{braket}
\usepackage{varioref}
\usepackage{xr-hyper}
\usepackage{xcolor}
\usepackage{nicefrac}
\usepackage{xfrac}
\usepackage{hyperref}
\hypersetup{colorlinks,linkcolor=blue,urlcolor=blue,citecolor=orange}
\usepackage{ulem}
\usepackage{siunitx}
\usepackage{graphicx}
\usepackage{dcolumn}
\usepackage{bm}
\usepackage{enumitem}

\usepackage{color}  

\newcommand{\blue}[1] {\textcolor{blue}{#1}}
\newcommand{\cut}[1] {\textcolor{green}{[----]}}

\begin{document}

\title{Stripe Symmetry of Short-range Charge Density Waves in Cuprate Superconductors}




\author{Jaewon Choi}
\thanks{These authors contribute equally to this work.}
\affiliation{Diamond Light Source, Harwell Campus, Didcot OX11 0DE, United Kingdom}

\author{Jiemin Li}
\thanks{These authors contribute equally to this work.}
\affiliation{Diamond Light Source, Harwell Science and Innovation Campus, Didcot OX11 0DE, United Kingdom}
\affiliation{Beijing National Laboratory for Condensed Matter Physics and Institute of Physics, Chinese Academy of Science, Beijing 100190, China}

\author{Abhishek Nag}
\affiliation{Diamond Light Source, Harwell Science and Innovation Campus, Didcot OX11 0DE, United Kingdom}

\author{Jonathan Pelliciari}
\affiliation{Department of Physics, Massachusetts Institute of Technology, Cambridge, Massachusetts 02139, USA}
\affiliation{National Synchrotron Light Source II, Brookhaven National Laboratory, Upton, New York 11973, USA}

\author{Hannah Robarts}
\affiliation{Diamond Light Source, Harwell Science and Innovation Campus, Didcot OX11 0DE, United Kingdom}
\affiliation{H. H. Wills Physics Laboratory, University of Bristol, Bristol BS8 1TL, United Kingdom}

\author{Charles C. Tam}
\affiliation{Diamond Light Source, Harwell Science and Innovation Campus, Didcot OX11 0DE, United Kingdom}
\affiliation{H. H. Wills Physics Laboratory, University of Bristol, Bristol BS8 1TL, United Kingdom}

\author{Andrew Walters}
\affiliation{Diamond Light Source, Harwell Science and Innovation Campus, Didcot OX11 0DE, United Kingdom}

\author{Stefano Agrestini}
\affiliation{Diamond Light Source, Harwell Science and Innovation Campus, Didcot OX11 0DE, United Kingdom}

\author{Mirian Garc\'{i}a-Fern\'{a}ndez}
\affiliation{Diamond Light Source, Harwell Science and Innovation Campus, Didcot OX11 0DE, United Kingdom}

\author{Dongjoon Song}
\affiliation{National Institute of Advanced Industrial Science and Technology (AIST), Tsukuba, Ibaraki 305-8560, Japan}

\author{Hiroshi Eisaki}
\affiliation{National Institute of Advanced Industrial Science and Technology (AIST), Tsukuba, Ibaraki 305-8560, Japan}

\author{Steven Johnston}
\affiliation{Department of Physics and Astronomy, The University of Tennessee, Knoxville, Tennessee 37996, USA}
\affiliation{Institute for Advanced Materials and Manufacturing, The University of Tennessee, Knoxville, Tennessee 37996, USA}

\author{Riccardo Comin}
\affiliation{Department of Physics, Massachusetts Institute of Technology, Cambridge, Massachusetts 02139, USA}

\author{Hong Ding}
\affiliation{Beijing National Laboratory for Condensed Matter Physics and Institute of Physics, Chinese Academy of Science, Beijing 100190, China}

\author{Ke-Jin Zhou}
\email{kejin.zhou@diamond.ac.uk}
\affiliation{Diamond Light Source, Harwell Science and Innovation Campus, Didcot OX11 0DE, United Kingdom}
\date{\today}

\begin{abstract}

The omnipresence of charge density waves (CDWs) across almost all cuprate families underpins a common organizing principle. However, a longstanding debate of whether its spatial symmetry is stripe or checkerboard remains unresolved. While CDWs in lanthanum- and yttrium-based cuprates possess a stripe symmetry, distinguishing these two scenarios has been challenging for the short-range CDW in bismuth-based cuprates. Here, we employed high-resolution resonant inelastic x-ray scattering to uncover the spatial symmetry of the CDW in Bi$_2$Sr$_{2-x}$La$_{x}$CuO$_{6+\delta}$. Across a wide range of doping and temperature, anisotropic CDW peaks with elliptical shapes were found in reciprocal space. Based on Fourier transform analysis of real-space models, we interpret the results as evidence of unidirectional charge stripes, hosted by mutually 90$^\circ$-rotated anisotropic domains. Our work paves the way for a unified symmetry and microscopic description of CDW order in cuprates.

\end{abstract}


\maketitle

\textbf{INTRODUCTION}\\
\\
High-temperature copper-oxide (or ``cuprate”) superconductors are prime examples of quantum materials that exhibit a unique set of broken-symmetry phenomena, such as pseudogap, strange metal behavior, nematicity, and various density-wave states \cite{Keimer2015,Norman2011}. Among these, charge-density waves (CDWs) --- periodic modulations of valence electron density --- have been ubiquitously discovered in both hole- and electron-doped cuprates \cite{Tranquada95,Tranquada96,Zimmermann1998,Fujita2002,Abbamonte2005,Wu2011,Ghiringhelli2012,Chang2012,Hanaguri2004,Hoffman2002,Howald2003,Wise2008,Vershinin2014,Neto2014,Cai2016,Li2021,Comin2014,Neto2015,Neto2016}. This universality has prompted suggestions of a unified principle underlying their formation.

\begin{figure*}[t]
\center{\includegraphics[width=0.92\textwidth]{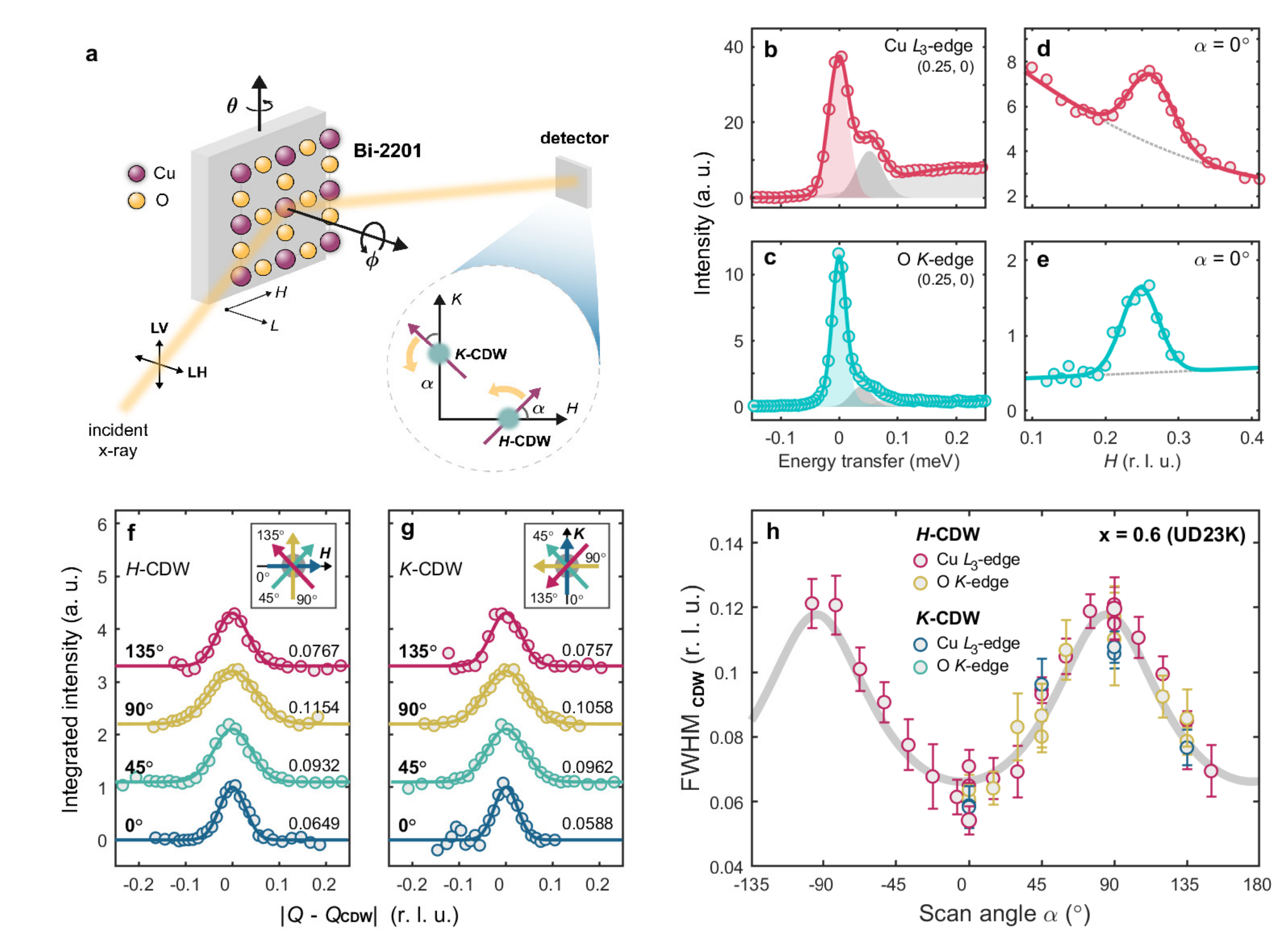}}
\caption{\textbf{Angle-dependent RIXS experiment on charge-density wave (CDW) in La-Bi2201.} \textbf{a}, Schematic illustration of experiment. By changing an in-plane sample rotation $\phi$, $H$- and $K$-CDWs are sliced along different directions, defined by azimuthal angle $\alpha$, in ($H$, $K$) plane. \textbf{b},\textbf{c}, Representative RIXS spectra at $\textbf{Q}=(0.25, 0)$ with Cu $L_3$ (\textbf{b}) and O $K$ resonance (\textbf{c}). Spectral weight from quasielastic, phonon, and paramagnon contribution is highlighted by red or cyan, dark-grey and light-grey color, respectively. \textbf{d},\textbf{e}, Integrated intensity of quasielastic peaks extracted from RIXS spectra as a function of $H$ at Cu $L_3$- and O $K$-edge. Solid lines are Gaussian fits with polynomial background. \textbf{f},\textbf{g}, Background-subtracted momentum-dependent profiles of integrated intensity across $H$-CDW (\textbf{f}) and $K$-CDW (\textbf{g}) at Cu $L_3$-edge with different scan angle $\alpha$. The insets depict scan directions. The number on the right side of each profile corresponds to fitted full-width half-maximum (FWHM). \textbf{h}, FWHMs of $H$- and $K$-CDW peaks measured at both absorption edges are plotted as a function of $\alpha$. The grey solid line indicates the least square fit to elliptical equation (See Methods). All data were obtained at $T=20$ K.
}	
\label{fig1}
\end{figure*} 

However there has been a longstanding controversy, that is, whether its spatial symmetry, $i.e.$, the spatial distribution of the CDW ordered state, is bidirectional ``checkerboard" or unidirectional ``stripe" \cite{Comin2016}. Although the presence of intertwined spin-charge stripe order is widely perceived, especially in lanthanum (La)-based cuprates \cite{Tranquada96}, the formation of mutually 90$^\circ$-rotated charge stripe domains due to two equivalent Cu-O bond directions in square-lattice CuO$_2$ plane leads to experimental results virtually indistinguishable from those of checkerboard CDW. For example, both CDW symmetries commonly generate four-fold electronic diffraction peaks along $H$ and $K$ directions in reciprocal space \cite{Comin2015,Comin2015NatMat}. Recent resonant x-ray scattering (RXS) experiments employed uniaxial pressure as an external symmetry-breaking perturbation to lift the degeneracy of two symmetry candidates, and successfully revealed a unidirectional nature of the CDW in La- \cite{Choi2022,Simutis2022} and yttrium (Y)-based cuprates \cite{Kim2018,Kim2021}. For bismuth (Bi)-based compounds, one of three prototypical material classes in cuprates, the CDW symmetry remains elusive. A plethora of scanning tunnelling microscopy (STM) studies reported a short-range checkerboard-like electronic modulation along Cu-O bond directions in real space with a four-unit-cell periodicity \cite{Hoffman2002,Howald2003,Wise2008,Vershinin2014,Neto2014,Cai2016,Li2021}. Interestingly, several STM studies revealed unidirectional nano-domains of electronic modulation breaking fourfold rotational symmetry \cite{Kohsaka2007,Lawler2010,Kohsaka2012}.  

Elucidating the CDW symmetry in Bi-based cuprates will therefore be indispensable for answering the question: is the unidirectional ``stripe" CDW symmetry universal for all different cuprate materials? Since the CDW is intimately related with superconductivity, this question may be essential for understanding the interplay of CDW, SC and other broken-symmetry states. For example, there have been a number of theoretical works suggesting that the fluctuating stripe may play a significant role in the mechanism of SC \cite{Kivelson1998,Kivelson2003,Cvetkovic2007,Huang2017,Mai2022}. In addition, a unidirectional pair-density-wave order (PDW), where a SC order parameter is intertwined with CDW and periodically modulated, has been proposed as a ``primary" order in cuprates \cite{Fradkin2010}. The exact answer to the question of CDW spatial symmetry is pivotal to differentiate microscopic models of PDW order parameter \cite{Agterberg2020}.

\begin{figure*}[t]
\center{\includegraphics[width=0.93\textwidth]{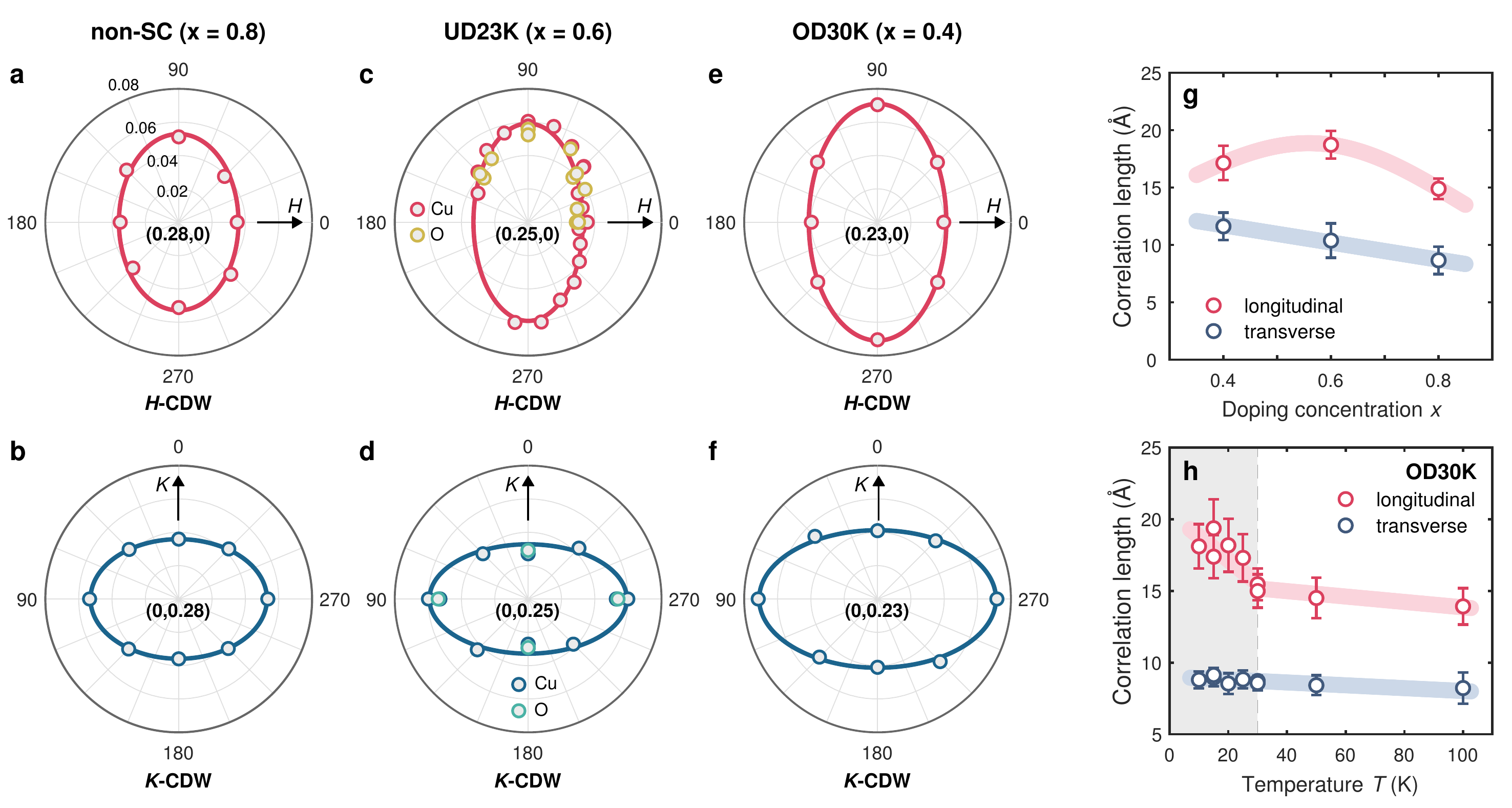}}
\caption{\textbf{Doping and temperature evolution of anisotropic CDW correlation.} \textbf{a}-\textbf{f}, Full-width half-maximum of both $H$-CDW (upper panels) and $K$-CDW (lower panels) peaks in La-Bi2201 with different doping concentrations as stated. Solid lines are least square fits to elliptical equation (see Methods). \textbf{g,h}, Doping concentration $x$ (\textbf{g}) and temperature $T$ dependence (\textbf{h}) of longitudinal and transverse CDW correlations. The grey dashed line indicates the superconducting transition temperature $T_c$. The colored trends are guided to the eye.
}	
\label{fig2}
\end{figure*} 
To address this question, we performed high-resolution resonant inelastic x-ray scattering (RIXS) measurements on Bi$_2$Sr$_{2-x}$La$_{x}$CuO$_{6+\delta}$ (La-Bi2201) to map out the CDW patterns in a broad range of reciprocal space. Our systematic studies reveal that the CDW patterns in La-Bi2201 possess an elliptical shape, elongated along the transverse direction. This anisotropic CDW pattern was robustly found in a wide range of temperatures and doping concentrations. By Fourier-transform analysis of real-space models, we concluded that equally populated 90$^\circ$-rotated domains of charge stripe orders lead to anisotropic CDW peaks. Our results not only provide evidence for the unidirectional stripe symmetry in Bi-based cuprates, but also suggest that the stripe CDW symmetry is another universal property shared among different cuprate material families. \\

\textbf{RESULTS}\\
\\
\textbf{Anisotropic CDW peaks} \\
The CDW order in La-Bi2201 system manifests itself as quasielastic scattering intensity peaks at four symmetry-equivalent positions in reciprocal space, ($H$, $K$) = ($\pm\delta$, 0) and (0, $\pm\delta$), where the incommensurability $\delta\sim{1/4}$ reciprocal lattice unit (r.l.u.) is consistent with the previous reports \cite{Neto2014,Li2020}. By rotating the in-plane azimuthal angle $\phi$ with respect to the CDW peak, we obtained momentum-dependent RIXS spectra across the CDW reflections at ($\delta$, 0) and (0, $\delta$), hereafter denoted as $H$- and $K$-CDW, with different scan angle $\alpha$ as illustrated in Fig.~\ref{fig1}\blue{a}. Here, $\alpha$ is defined as the angle between the actual scan and the longitudinal propagating direction. The incoming photon energy was tuned to the resonances of Cu $L_3$-edge ($\sim$931.6 eV) and O $K$-edge ($\sim$528.4 eV) absorption energy to maximize the sensitivity to CDW. Energy-resolved high-resolution RIXS has an advantage of disentangling a weak quasielastic short-range CDW response from the phonon modes as well as the spin and orbital excitations (Fig.~\ref{fig1}\blue{b,c}), comparing to the energy-integrated RXS \cite{Comin2014}. An enhancement in the integrated quasielastic scattering intensity is clearly identified near $\textbf{Q}=(0.25, 0)$ and $(0, 0.25)$ at both Cu and O sites, confirming its multiorbital nature \cite{Li2020} (Fig.~\ref{fig1}\blue{d,e}). 


\begin{figure*}[t]
\center{\includegraphics[width=0.93\textwidth]{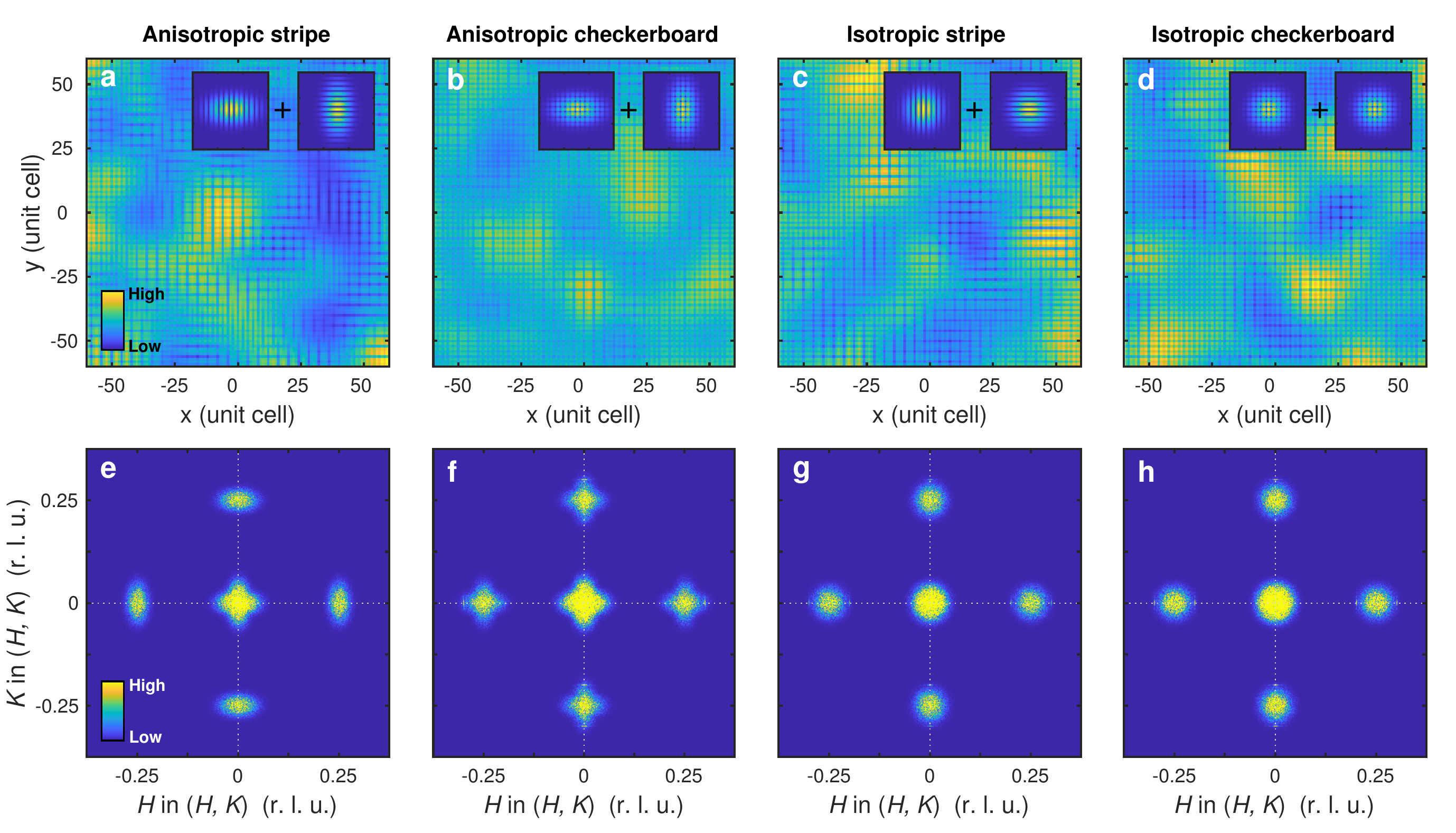}}
\caption{\textbf{Model CDW modulation in real and reciprocal space.} \textbf{a}-\textbf{d}, Real-space electron density modulation map $\rho(x,y)$ calculated by assuming equal population of (\textbf{a}) anisotropic 90$^\circ$-rotated stripe domains, (\textbf{b}) 90$^\circ$-rotated anisotropic checkerboard domains, (\textbf{c}) 90$^\circ$-rotated isotropic stripe domains, and (\textbf{d}) isotropic checkerboard domains. \textbf{e}-\textbf{h}, Corresponding CDW peak patterns in reciprocal space, obtained by direct Fourier transformation of (\textbf{a}-\textbf{d}), respectively.
}	
\label{fig3}
\end{figure*} 


To gain an overview of CDW patterns, we show representative CDW scans obtained at Cu $L_3$-edge at selected values of $\alpha$ across $H$-CDW (Fig.~\ref{fig1}\blue{f}) and $K$-CDW (Fig.~\ref{fig1}\blue{g}) of underdoped La-Bi2201 ($x=0.6$, UD23K). The full-width half-maximum (FWHM) of $H$-CDW in the transverse direction ($\alpha=90^\circ$) is noticeably larger than that in the longitudinal direction ($\alpha=0^\circ$), demonstrating an anisotropic CDW peak structure (Fig.~\ref{fig1}\blue{f}). Examination of $K$-CDW reveals the same anisotropy with comparable FWHMs (Fig.~\ref{fig1}\blue{g}). We also found the elongated CDWs present along both $H$ and $K$ directions at O $K$-edge (see Supplementary Information). 


We now scrutinize the analysis on the shape of the individual CDW peaks. Fig.~\ref{fig1}\blue{h} summarises the extracted FWHM of $H$-CDW from both Cu $L_3$- and O $K$-edges as well as those of $K$-CDW as a function of $\alpha$ from -97.5$^\circ$ to 150$^\circ$ (Supplementary Figs.~\blue{2} and \blue{3}). Clearly the data show an oscillatory behavior as a function of $\alpha$. The least square fit of the data to an elliptical equation (the grey solid line in Fig.~\ref{fig1}\blue{h}, Methods) returns the maximum $\sim0.118$ r.l.u. near $\alpha=90^\circ$ and the minimum $\sim0.066$ r.l.u. near $\alpha=0^\circ$. The major and minor axes along 0$^\circ$ and 90$^\circ$ indicate that the CDW propagation is locked along two orthogonal Cu-O bond directions. The aspect ratio of 1.8 implies the CDW correlation, $i.e.$, $\xi=2/\textrm{FWHM}$, in the longitudinal direction ($\xi_\parallel\sim$18.7~\AA) is almost twice as long as that in the transverse direction ($\xi_\perp\sim$10.4~\AA). In Fig.~\ref{fig2}\blue{c,d}, the same set of data are plotted in polar coordinates. The shape of the individual CDW peak is best described as an ellipse elongated along the transverse direction, perpendicular to the direction of CDW propagation. The $K$-CDW ellipse possesses almost the same size but rotated by $90^\circ$, implying that the CDW pattern in reciprocal space preserves fourfold ($C_\textrm{4}$) symmetry, seemingly reflecting the nature of the underlying CuO$_2$ square-lattice.\\
\\
\textbf{Temperature and doping evolution} \\
We extended the same RIXS study to explore the doping and temperature dependence. Fig.~\ref{fig2}\blue{a,b}, and Fig.~\ref{fig2}\blue{e,f} exhibit the CDW peaks obtained from a non-superconducting ($x=0.8$, non-SC) and an optimally-doped ($x=0.4$, OD30K) La-Bi2201, respectively. Fig.~\ref{fig2}\blue{g} summarises the longitudinal correlation length, $\xi_\parallel$, and the transverse correlation length, $\xi_\perp$, of the samples with different doping concentration $x$. $\xi_\parallel(x)$ reaches its maximum at $x=0.6$. On the other hand, $\xi_\perp(x)$ shows a monotonically decreasing trend. The temperature dependence of $\xi_\parallel$ and $\xi_\perp$ in OD30K sample are plotted in Fig.~\ref{fig2}\blue{h}. $\xi_\parallel$ increases much faster below $T_c$, whereas $\xi_\perp$ displays a continuous linear trend. These results confirm the robust existence of the elongated CDW in a wide range of temperature and doping concentration.\\
\\
\textbf{Modelling CDW patterns} \\
The CDW peak patterns in reciprocal space probed by RIXS encode the spatial symmetry of the CDW order in real space. We thus performed theoretical simulations by constructing idealized models, that is, the real-space electronic modulation with either stripe or checkerboard pattern, where both anisotropic and isotropic domains are considered. For each scenario, two individual CDW domains are included with a $90^\circ$ rotation from each other to mimic the equally populated orthogonal domains (insets of Fig.~\ref{fig3}\blue{a-d}, see Supplementary Information). The real-space electron density modulation maps $\rho(x,y)$ were built by translating the orthogonal CDW domains in the x and y directions with a random phase shift to represent the actual samples probed by RIXS (Fig.~\ref{fig3}\blue{a-d}). Fourier transforms of the electron density maps $\rho(x,y)$ yield a unique CDW reciprocal-space pattern for each scenario (Fig.~\ref{fig3}\blue{e-h}). For instance, an anisotropic stripe symmetry produces elliptical CDW diffraction peaks, preserving the C$_4$ symmetry but elongated in the transverse direction (Fig.~\ref{fig3}\blue{e}). The anisotropic checkerboard domains give rise to star-shaped CDW peaks elongated along both the transverse and longitudinal directions (Fig.~\ref{fig3}\blue{f}). On the other hand, isotropic checkerboard and isotropic stripe phases generate virtually indistinguishable isotropic CDW peaks (Fig.~\ref{fig3}\blue{g,h}). Remarkably, our experimental observations in La-Bi2201 (Fig.~\ref{fig2}) coincide with the spatial symmetry of CDW order having unidirectional stripe, rather than bidirectional checkerboard. A consistent conclusion is achieved based on similar Fourier transform analysis on more realistic electron density maps (Supplementary Fig.~\blue{6})~\cite{Fine2016}.\\
\\

\textbf{DISCUSSION}\\
\\
At the first glance, it may look perplexing that we straightforwardly obtain the unambiguous agreement between experiment and theoretical modeling, while the CDW symmetry is still debatable despite a considerable number of STM studies \cite{Hoffman2002,Howald2003,Hanaguri2004,Wise2008,Vershinin2014}. The static checkerboard-like modulation was proposed for Bi-based cuprates from tunneling electron conductance maps. In their Fourier-transformed images, fourfold symmetry-equivalent peaks typically appear at an incommensurate wavevector position \cite{Hoffman2002,Howald2003,Wise2008,Vershinin2014,Neto2014,Cai2016,Li2021}. However, the conservation of $C_4$ symmetry is not a unique signature of the checkerboard-like modulation pattern as the coexistence of orthogonal domains of unidirectional stripe-like CDW also produces fourfold peak pattern. More sophisticated analysis on tunneling asymmetry at the pseudogap energy scale reveals the presence of four-unit-cell-wide unidirectional nanodomains breaking $C_4$ rotational symmetry \cite{Kohsaka2007,Lawler2010,Kohsaka2012}. This observation can be interpreted as a fluctuating ``stripe" with short correlation. It is worth noting that the nature of short-range CDW order imposes severe challenges on the assessment of the symmetry based on real-space images which manifest visually checkerboard-like CDW patterns regardless of whether its intrinsic symmetry is stripe or checkerboard, as shown by our modelled density maps (Fig.~\ref{fig3}\blue{a-d}) as well as advanced theoretical studies \cite{Robertson2006}. In this sense, RIXS is more advantageous as it probes the CDW spatial symmetry directly in reciprocal space, despite the short-range nature of the CDW order. 

Elongated CDW peaks in reciprocal space were also found in Y-based cuprates by RXS and proposed as evidences of unidirectional stripe-like CDW order \cite{Comin2015}. The interpretation was challenged as the corresponding real-space correlation function $C(x,y)$, $i.e.$, the direct Fourier transform of the reciprocal-space CDW pattern, similar to Fig.~\ref{fig3}\blue{e}, appears checkerboard-like \cite{Fine2016}. The calculation of $C(x,y)$ via auto-correlation of the real-space electron density maps (Fig.~\ref{fig3}\blue{a-d}) confirmed all four scenarios induce comparable checkerboard-like patterns akin to previous reports (Supplementary Fig.~\blue{5}) \cite{Fine2016}. Therefore, it is challenging to directly visualize the symmetry of a short-range CDW from either the electron density map $\rho(x,y)$  or correlation function $C(x,y)$ in real space. Instead of examining the shape of CDW peaks, a recent RIXS study under uniaxial pressure reveals a unidirectional nature of CDW in Y-based cuprates \cite{Kim2021}. The anisotropic CDW peaks we observed in La-Bi2201 are in stark contrast to isotropic CDW peaks reported in La-based \cite{Thampy2013,Thampy2014,Wang2020,Wang2021} and Hg-based cuprate materials \cite{Campi2015}. For La$_{1.88}$Sr$_{0.12}$CuO$_{4}$, twinned domains of slanted stripe order with $\pm3^\circ$ away from Cu-O bond directions result in two CDW peaks along transverse direction, eventually merging into a single isotropic peak under mild uniaxial pressure  \cite{Wang2021}. Despite the isotropic peak shape, x-ray diffraction employing uniaxial pressure unambiguously verifies charge stripe order \cite{Choi2022}.

Alongside the results from La- and Y-based cuprates, our RIXS measurements on La-Bi2201 corroborates  unidirectional stripes as the fundamental symmetry of CDW across different cuprate families. This remarkable universality not only provides a perspective on the unified organizing principle of CDW and its role in cuprate phase diagram, but also constrains the symmetry in microscopic models describing CDW and the interplay with other phases.\\
\\

\textbf{METHODS}\\
\\
\textbf{Sample growth and characterization} \\
High-quality single crystals of Bi$_2$Sr$_{2-x}$La$_{x}$CuO$_{6+\delta}$ (La-Bi2201) with hole-doping concentrations of $x = 0.8$ (non-SC), $x = 0.6$ ($T_{c}=23$ K, UD23K), $x = 0.4$ ($T_{c}=30$ K, OD30K) were synthesized by the travelling-solvent floating-zone method. The as-grown samples were annealed at 650$^\circ$C in an oxygen atmosphere for 2 days to improve the homogeneity of the samples. The sample orientation was characterized by Laue diffraction prior our resonant x-ray inelastic scattering (RIXS) experiments. Superstructure reflections propagating along Cu-Cu bond direction, or equivalently along ($H$, $H$) direction, were identified by the Laue diffraction method. All samples were cleaved by mechanical exfoliation, before transferring to a loadlock chamber. \\
\\
\textbf{X-ray absorption spectroscopy} \\
X-ray absorption spectra of La-Bi2201 samples were collected with a normal incidence geometry ($\theta = 90^\circ$) prior to the RIXS experiments. The total electron yield method was employed by measuring the drain current from the samples. Supplementary Figs.~\blue{1a,b} show representative x-ray absorption spectra of La-Bi2201 $x=0.6$ (UD23K) near Cu $L_3$ and O $K$ absorption edges, respectively. The data exhibit a single resonance peak at an energy of 931.6 eV and 528.4 eV. \\
\\
\textbf{High-resolution RIXS experiments}\\
We performed high-resolution RIXS experiments on La-Bi2201 at the I21 RIXS beamline of Diamond Light Source in the United Kingdom \cite{KJZhou2022}. The experimental geometry is schematically illustrated in Fig.~\ref{fig1}\blue{a} of the main text. Samples were glued on the holder such that its surface normal is the crystallographic $c$-axis direction. We define a wavevector $\textbf{Q}$ employing pseudo-tetragonal notation $\textbf{Q}=Ha^{*}+Kb^{*}+Lc^{*}$, or equivalently, $\textbf{Q}=(H, K, L)$ where $a^{*}=2\pi/a$, $b^{*}=2\pi/b$, $c^{*}=2\pi/c$ are the reciprocal lattice unit vectors and $H$, $K$, $L$ are the indices in reciprocal space spanned by $a^{*}$, $b^{*}$, and $c^{*}$. The lattice parameters of La-Bi2201 are $a=b=3.86$ \AA~and $c=24.69$ \AA. The incoming x-rays were tuned to the resonant energy of Cu $L_3$ and O $K$-edge to maximize the signal sensitivity to CDW residing in the CuO$_2$ plane. Linear vertical, or $\sigma$ polarization was used throughout the  measurements. The outgoing x-rays emitted from the La-Bi2201 sample pass through paraboloidal collecting optics with a large horizontal acceptance angle inside the main sample chamber to enhance the x-ray photon throughput. The spherical grating optics implemented on a spectrometer to vertically monochromatise emitted x-rays and focus to an area detector. The energy resolution of 37.0 meV and 27.3 meV was achieved for Cu $L_3$- and O $K$-edge, respectively. The scattering angle $2\theta$ is fixed to 154$^\circ$.\\
\\[4mm]
\textbf{RIXS data fitting} \\
RIXS spectra were firstly normalized to the incident photon flux and the spectral weight of \textit{dd} excitation integrated over an energy range of 1.2-3 eV. The position of the quasi-elastic peak, or zero-energy position, was determined by measuring the amorphous carbon sample. Compared to energy-integrated resonant elastic x-ray scattering (REXS), the energy-resolved RIXS has the advantage of filtering out unwanted contributions from inelastic scatterings due to various low-energy excitations such as phonon, magnon, and orbital (\textit{dd}) excitations. To better disentangle quasielastic diffraction intensity, the low-energy region of RIXS spectra measured at the Cu $L_3$-edge was fitted with a sum of quasielastic, phonon, paramagnon, and background contributions as follows:
\begin{equation} \label{eqn:1}
I_{RIXS}^{Cu}(E)=I_{el}^{Cu}(E)+I_{ph}^{Cu}(E)+I_{mag}^{Cu}(E)+I_{bg}^{Cu}(E).
\end{equation}
Similarly for the O $K$-edge, 
\begin{equation} \label{eqn:2}
I_{RIXS}^{O}(E)=I_{el}^{O}(E)+I_{ph}^{O}(E)+I_{bg}^{O}(E).
\end{equation}
Four terms on the right-hand side of Eq. (\ref{eqn:1}) represent a quasielastic, phonon, paramagnon, and background contribution on the RIXS intensity, respectively. The quasielastic peak and phonon excitation are described by Gaussian functions, whereas the paramagnon excitation is modelled by the response function of a damped harmonic oscillator multiplied with the Bose factor \cite{Monney2016}, as follows:
\begin{gather}
I_{el}^{Cu,O}(E)=A_1exp\bigg\{{\frac{1}{2}\Big(\frac{E-A_{2}}{A_{3}}\Big)^2}\bigg\} \label{eqn:2}\\
I_{ph}^{Cu,O}(E)=\sum_{i=1}^{n}B_{3i-2}exp\bigg\{{\frac{1}{2}\Big(\frac{E-B_{3i-1}}{B_{3i}}\Big)^2}\bigg\} \label{eqn:3}\\
I_{mag}^{Cu}(E)=C_1\bigg\{\frac{C_3E}{(E^2-C_2^2)^2+E^2C_3^2}\bigg\}\times\bigg\{\frac{1}{1-e^{-E/C_4}}\bigg\} \label{eqn:4}\\[3mm]
I_{bg}^{Cu,O}(E)=D_1E^2+D_2E+D_3 \label{eqn:5}
\end{gather}
An integer $n$ indicates the number of phonon modes: $n=1$ (only bond-stretching mode) for the Cu $L_3$-edge and $n=2$ (bond-stretching and bond-buckling modes) for the O $K$-edge \cite{Li2020}. $A_1, A_2, A_3$, $B_1, \cdots ,B_n$, and $C_1, C_2, C_3$ are the fitting parameters. $C_4$ was fixed to $k_{B}T$ with the measuring temperature $T$.  \\
\\
\textbf{Angular-dependent CDW measurement} \\
Taking advantage of broad accessibility to in-plane rotation angle and accurate control of sample manipulator, we obtained angle- and momentum-dependent RIXS spectra across both $H$- and $K$-CDWs. By controlling the polar angle $\theta$, the azimuthal rotation angle $\phi$, and the tilt angle $\chi$ using diffcalc, momentum-dependent RIXS scans can be obtained across the CDW peak with different scan angle $\alpha$. Here, the scan angle $\alpha$ is defined as an angle between the scan direction and the longitudinal scan direction for each CDW. The representative scan directions are schematically described in the inset of Figs.~\blue{1f,g}. For Cu $L_3$-edge, the CDW peaks were measured in a range between -97.5$^\circ$ and 150$^\circ$. For O $K$-edge, $\alpha$ was varied from 0$^\circ$ to 135$^\circ$. Supplementary Figs.~\blue{2} and \blue{3} present the momentum-dependence of integrated quasielastic intensity scans on $H$-CDW taken at different $\alpha$ as indicated. The momentum-dependence data is fitted with a combination of Gaussian functions, which represents a CDW peak, and a composite background.
\begin{equation}
\begin{split}
    I_{CDW}(Q)=G_1exp\bigg\{{\frac{1}{2}\Big(\frac{Q-G_{2}}{G_{3}}\Big)^2}\bigg\}+G_{5}exp\big(\frac{x-G_{6}}{G_{7}}\big)\\
    +G_{8}Q^2+G_{9}Q+G_{10}.
\end{split}
\end{equation}
A full-width half-maximum (FWHM) of CDW peak is calculated by $2\sqrt{2\textrm{ln}2}G_3$ for each momentum-dependent profile. In Supplementary Figs.~\blue{2} and \blue{3}, the least-square fitted results are shown with solid lines at different scan angle $\alpha$. The FWHMs of CDW peak are written in each panel. The angle-dependent FWHM shown in Fig.~\ref{fig1}\blue{h} is fitted with an elliptical equation stated below:
\begin{equation}
    FWHM(\alpha)=\frac{R_1R_2}{\sqrt{\big\{bcos(\alpha+\beta)\big\}^2+\big\{bsin(\alpha+\beta)\big\}^2}}
\end{equation}

where $R_1$ ($R_2$) is the major (minor) axis of the ellipse and $\beta$ is an offset angle. The grey-coloured line presented in Fig.~\ref{fig1}\blue{h} is the result of least square fitting which gives $R_1 = 0.1184\pm0.0022$ r.l.u., $R_2 = 0.0656\pm0.0013$ r.l.u., and $\beta=1.2\pm1.9^\circ$. The offset angle $\beta\sim0$ indicates that the CDW propagation is locked along two mutually-orthogonal Cu-O bond directions. For La-Bi2201 $x=0.6$ (UD23K) samples, the data were obtained for three sample pieces from different batches.\\
\\
\textbf{Acknowledgements}\\
We thank Johan Chang, Stephen M. Hayden, Wei-Sheng Lee, and Qisi Wang for fruitful discussions. We also acknowledge Thomas Rice, the beamline technician at I21 beamline, for his technical supports. We thank Gavin Stenning and Daniel Nye for the Laue diffraction in the Materials Characterisation Laboratory at the ISIS Neutron and Muon Source. This work was primarily supported by in-house research program of Diamond Light Source, Ltd. through beamtime proposals NT21277 and NT29623. J. L. acknowledges Diamond Light Source, Ltd. and Institute of Physics in Chinese Academy of Science for providing funding Grant 112111KYSB20170059 for the joint Doctoral Training under Contact STU0171. H. R. acknowledges funding and support from Engineering and Physical Sciences Research Council Centre for Doctoral Training through Grant EP/L015544/01 and EP/R0111141/1. H. D. acknowledges the financial support from the National Science Foundation of China (Grant 11888101), and the Ministry of Science and Technology of China (Grant 2016YFA0401000). S. J. acknowledges support from NSF under Grant DMR-1842056. J. P. acknowledges financial support by the SNSF Postdoc Mobility Project P400P2\textunderscore180744.\\ 
\\
\textbf{Author contributions}\\
K. Z. conceived and supervised the project. J. C., J. L., A. N., J. P., H. R., C. C. T., A. W., S. A., M. G.-F. and K. Z. performed RIXS experiment. D. S. and H. E. synthesized and prepared the materials. J. C., J. L., and K. Z. analyzed the data. J. C. and K. Z. wrote the manuscript with input from all coauthors.\\
\\
\textbf{Competing interests}\\
The authors declare no competing interests.\\
\\
\textbf{Additional information}\\
Supplementary information is available for this paper. Correspondence and requests for materials should be addressed to Jaewon Choi and Ke-Jin Zhou.

\bibliography{bi-2201_ref}

\newcommand{\beginsupplement}{
        \setcounter{table}{0}
        \renewcommand{\thetable}{S\arabic{table}}
        \setcounter{figure}{0}
        \renewcommand{\figurename}{\textbf{Supplementary Fig.}}}
\beginsupplement

\onecolumngrid

\newpage

\section{Modeling CDW - simplified cases}

To understand the physical implication of CDW patterns in reciprocal space, we constructed electronic density maps using simplified real-space models by taking into account the following assumptions based on experimental observations:  
\begin{itemize}
    \item The spatial symmetry of CDW is either stripe or checkerboard. The CDW propagates along two Cu-O bond directions (either crystallographic a or b direction) with a four-unit-cell periodicity.
    \item The shape of domains hosting CDW is either isotropic or anisotropic. The size of individual domains is proportional to the CDW correlation.
    \item In the case of stripe CDW, two kinds of CDW domains propagating along either a or b direction are equally populated, as the system does not energetically favor one or another. We thus simulate the situation by having mutually 90$^\circ$ rotated anisotropic domains with equal populations.
    \item The real space electron density maps comprise domains distributed with a random phase shift.\\
\end{itemize}

\noindent\textbf{1. Individual CDW domains}

\noindent For the first step, a real-space electronic density modulation of individual CDW domains was modelled. Supplementary Fig.~\ref{fig:S4} shows individual CDW domains with (a,b) anisotropic horizontal and vertical stripe, (c,d) anisotropic horizontal and vertical checkerboard, (e,f) isotropic horizontal and vertical stripe, and (g) isotropic checkerboard modulations with 4-unit-cell periodicity, propagating along two orthogonal Cu-O bond directions. \\
\\
\textbf{2. Model Real-space intensity maps} \\
For the next step, an electronic density modulation map $\rho(x,y)$ (1000$\times$1000 unit cells) was modelled in real space by translating the individual CDW domains into $x$ and $y$ directions with a random phase shift. Model $\rho(x,y)$ employing individual domains as building blocks allows us to reveal how the CDWs with different spatial symmetry result in different macroscopic real-space and reciprocal-space intensity patterns. Supplementary Figs.~\ref{fig:S5}a-c represent electronic density map $\rho(x,y)$ in real space which consists equally populated horizontal and vertical domains which are mutually rotated by 90$^\circ$ from each other and are profiled with (a) anisotropic stripe, (b) anisotropic checkerboard, (c) isotropic  stripe, respectively. Supplementary Fig.~\ref{fig:S5}d corresponds to the case having isotropic checkerboard domains. The microscopic structure of individual domains is depicted in the insets for each panel. These maps do not necessarily illustrate an electronic modulation in a single layer, as random translation inevitably leads to overlaps of CDW domains. Considering a penetration depth $\lambda>100$ nm for the Cu $L_3$-edge and $\lambda>50$ nm for the O $K$-edge, RIXS measures an average of more than 40 (Cu) or 20 (O) unit cells stacked along $c$-axis. Therefore, the real-space density maps shown in Supplementary Fig.~\ref{fig:S5}a-d provide an excellent model of the electronic modulations in La-Bi2201 probed by RIXS technique. $\rho(x,y)$ in all four scenarios look very similar to each other. It is visually impossible to distinguish whether the intrinsic symmetry of individual CDW domains is stripe or checkerboard, solely based on $\rho(x,y)$. This difficulty had already been raised by Robertson \textit{et al.} (2006) \cite{Robertson2006}, especially in highly disordered systems hosting short-ranged CDW where its correlation length $\xi_\textrm{CDW}$ is comparable to the periodicity, such as in La-Bi2201.\\
\\
\textbf{3. Simulated CDW patterns } \\
Supplementary Figs.~\ref{fig:S5}e-h show intensity maps $S(H,K)$ in reciprocal space obtained by a direct Fourier transformation of $\rho(x,y)$ shown in Supplementary Figs.~\ref{fig:S5}a-d. The calculated $S(H,K)$ is equivalent to the structure factor measured by scattering experiments. $S(H,K)$ from anisotropic 90$^\circ$-rotated stripe domains gives four CDW peaks elongated along the transverse direction (Supplementary Figs.~\ref{fig:S5}a,e), which can be clearly differentiated from other three cases. Equally-populated anisotropic checkerboard domains result in four CDW peaks with a cross-like shape (Supplementary Figs.~\ref{fig:S5}b,f). Interestingly, $S(H,K)$ from both isotropic stripe and checkerboard domains are almost identical (Supplementary Figs.~\ref{fig:S5}c,d,g,h). This implies that it is difficult to differentiate the stripe and checkerboard modulation in reciprocal space, when the CDW domains are isotropic.  \\
\\
\textbf{4. Correlation function } \\
Fourier transforms of $S(H,K)$ give a density-density correlation function $C(x,y)\cong\langle\rho(x+\Delta x,y+\Delta y)\rho(x,y)\rangle$, which is mathematically equivalent to an auto-correlation of $\rho(x,y)$. Supplementary Figs.~\ref{fig:S5}i-l present $C(x,y)$ calculated from the auto-correlation of $\rho(x,y)$ shown in Supplementary Fig.~\ref{fig:S5}a-d. The results from all four symmetry scenarios provide qualitatively similar checkerboard-like pattern. This checkerboard-like $C(x,y)$ is consistent with Ref.~\cite{Fine2016}. We confirm that neither auto-correlation function $C(x,y)$, nor the electron density map in real space could provide visually unambiguous clarification on the symmetry being either stripe or checkerboard.

\section{Modeling CDW - more realistic cases}


In this section, we generate more realistic electronic density map $\rho(x,y)$ in which distortions of continuous CDW domains are considered to represent partially experimental observations by STM \cite{Hoffman2002,Howald2003,Wise2008,Vershinin2014,Neto2014,Cai2016,Li2021}. The structure factor $S^\textrm{CDW}(H, K)$ in the reciprocal space can be described as follows:
\begin{equation}
    S^\textrm{CDW}(H, K)=\sum_{i=1}^{4}A_i\textrm{exp}\bigg[-\bigg\{\frac{(H-Q_{x,i})^2}{2\sigma_{x,i}^2}+\frac{(K-Q_{y,i})^2}{2\sigma_{y,i}^2}\bigg\}\bigg],
\end{equation}
Here, $A_i$ and $\textbf{Q}_{i}$, represent the CDW order parameter and wavevector, respectively, are fixed to $A_i=1$, $\textbf{Q}_{i}=(Q_{x,i},Q_{y,i})=(\pm0.25, 0)$ or $(0, \pm0.25)$ commonly in all three scenarios. $\sigma_{x,i}$ and $\sigma_{y,i}$ are the width of four individual CDW peaks along the x and y directions, respectively. We constructed $S^\textrm{CDW}(H, K)$ for the following three different scenarios as shown in Supplementary Figs.~\ref{fig:S6}a-c: 
\begin{enumerate}[label=(\alph*)]
    \item Four anisotropic CDW peaks elongated along the transverse direction, where $\sigma_{y,i}=2\times\sigma_{x,i}$ for $\textbf{Q}_{i}=(\pm0.25, 0)$ and $\sigma_{x,i}=2\times\sigma_{y,i}$ for $\textbf{Q}_{i}=(0, \pm0.25)$. The calculated $S(H,K)$ is shown in Supplementary Fig.~\ref{fig:S6}a, which is similar to Supplementary Fig.~\ref{fig:S5}e, describing 90$^\circ$-rotated \textbf{anisotropic charge stripe domains}.
    \item Four isotropic CDW peaks, where $\sigma_{x,i}=\sigma_{y,i}$. The calculated $S(H,K)$ is shown in Supplementary Fig.~\ref{fig:S6}b, which is similar to Supplementary Figs.~\ref{fig:S5}g and~\ref{fig:S5}h, describing \textbf{isotropic stripe or checkerboard domains}.
    \item Four star-like CDW peaks, obtained by the average of $S_{c1}(H,K)$ and $S_{c2}(H,K)$. $S_{c1}(H,K)$ is identical to $S(H,K)$ in the scenario (a). $S_{c2}(H,K)$ represents CDW peaks elongated along the longitudinal direction where $\sigma_{x,i}=2\times\sigma_{y,i}$ for $\textbf{Q}_{i}=(\pm0.25, 0)$ and $\sigma_{y,i}=2\times\sigma_{x,i}$ for $\textbf{Q}_{i}=(0, \pm0.25)$. The calculated $S(H,K)$ is shown in Supplementary Fig.~\ref{fig:S6}c, which is similar to that in Supplementary Fig.~\ref{fig:S5}f, describing 90$^\circ$-rotated \textbf{anisotropic checkerboard domains}.
\end{enumerate}
Based on the structure factor $S(H, K)$ for the scenario (a)-(c) (Supplementary Figs.~\ref{fig:S6}a-c), we generated electron density maps $\rho(x,y)$ by a direct Fourier transformation of $S(H, K)$ employing the following method \cite{Fine2016}: 
\begin{equation}
    \rho(x,y)=\sum_{m}\sqrt{S(H_{m}, K_{m})}cos(H_{m}x+ K_{m}y+\phi)
\end{equation}
Here, $m$ is an index of each pixel in a discrete $1000\times1000$ grid of $S(H,K)$. $\phi$ is the random phase shift. The generated $\rho(x,y)$ map for each $S(H,K)$ is presented in Supplementary Figs.~\ref{fig:S6}d-f where only $100\times100$ unit-cell grid is shown for the sake of visibility. They surprisingly resemble the conductance map in real space obtained by STM  \cite{Hoffman2002,Howald2003,Wise2008,Vershinin2014,Neto2014,Cai2016,Li2021}. Similar to Supplementary Figs.~\ref{fig:S5}a-d, it is difficult to find qualitative differences among these three $\rho(x,y)$ maps. Despite added randomness, the Fourier-transformed $\rho(x,y)$ shown in Supplementary Figs.~\ref{fig:S6}g-i demonstrate that the symmetry of CDW peak patterns is better revealed in reciprocal space. These results are qualitatively consistent with those from our idealized models summarized in Section II.

\newpage

\begin{figure*}[t]
\center{\includegraphics[width=0.8\textwidth]{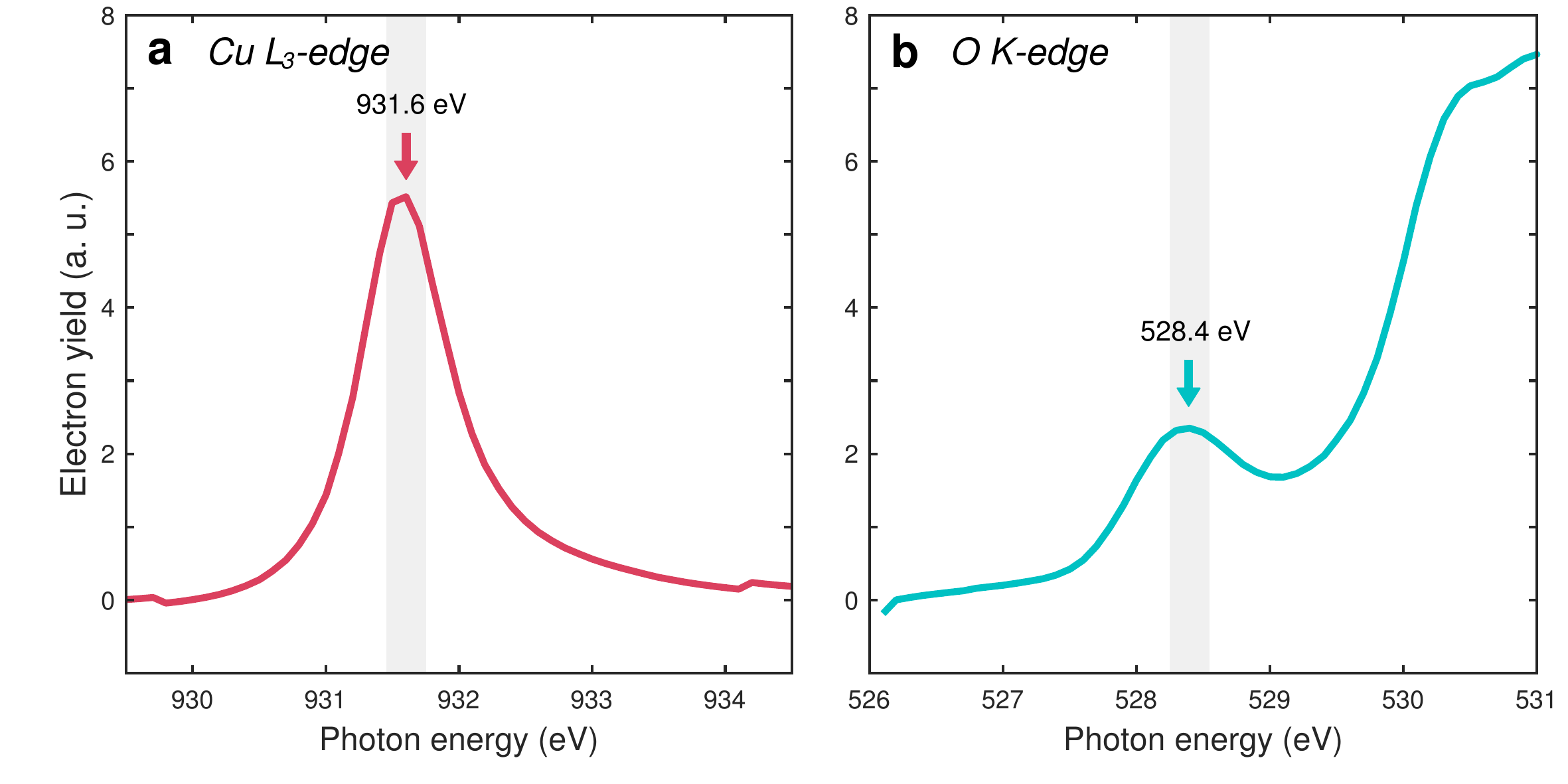}}
\caption{\textbf{X-ray absorption spectroscopy on Bi$_2$Sr$_{1.4}$La$_{0.6}$CuO$_{6+\delta}$.} \textbf{a},\textbf{b} Cu $L_3$-edge (\textbf{a}) and O $K$-edge (\textbf{b}) x-ray absorption spectroscopy (XAS) on Bi$_2$Sr$_{1.4}$La$_{0.6}$CuO$_{6+\delta}$ (La-Bi2201) sample collected with linear-vertical (LV) polarization of incoming x-ray with normal incident geometry ($\theta=90^\circ$).}	
\label{figs1}
\end{figure*} 

\newpage

\begin{figure*}[t]
\center{\includegraphics[width=0.98\textwidth]{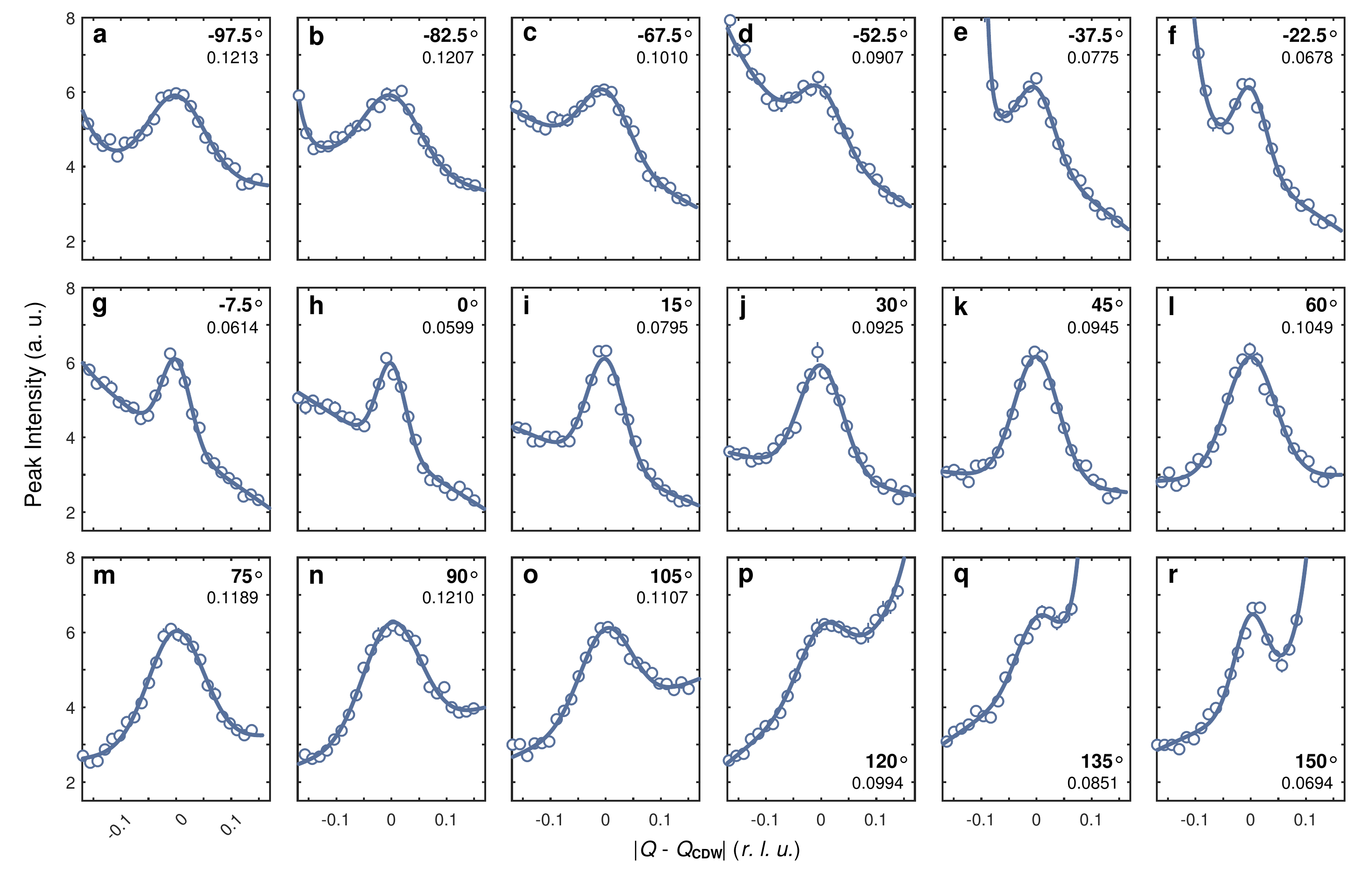}}
\caption{\textbf{Charge density wave measured at Cu \textit{L}$_3$-edge.} \textbf{a}-\textbf{r}, Momentum-dependent profile of integrated quasi-elastic charge-density-wave (CDW) peak intensity measured at different scan angle $\alpha$ as indicated with a bold text. The incident photon energy was fixed to the resonance of Cu $L_3$ absorption edge. The solid line is fit to Gaussian function representing the CDW peak with a composite background (See Methods). The full-width half-maximum (FWHM) of CDW peak is also presented on each panel.  }	
\label{fig:S2}
\end{figure*} 

\newpage

\begin{figure}[htb]
 	\begin{center}
 		\includegraphics[width=0.8\textwidth]{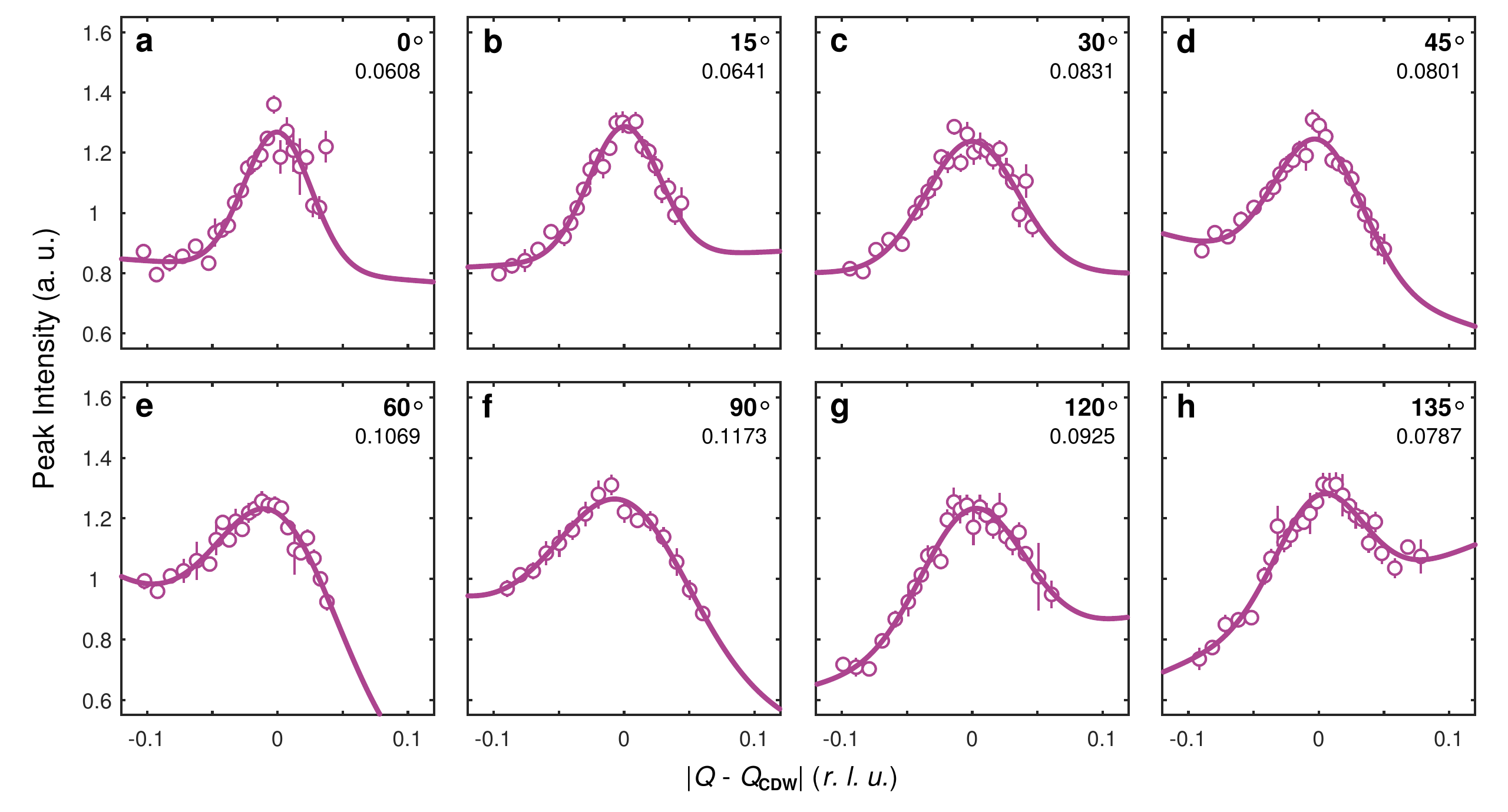}
 	\end{center}
 	\caption{\textbf{Charge density wave measured at O \textit{K}-edge.} \textbf{a-h}, Momentum-dependent profile of integrated quasi-elastic charge-density-wave (CDW) peak intensity measured at different scan angle $\alpha$ as indicated with a bold text. The incident photon energy was fixed to the resonance of O $K$ absorption edge. The solid line is fit to Gaussian function representing the CDW peak with a polynomial background. The full-width half-maximum (FWHM) of CDW peak is also presented on each panel.} \label{fig:S3}
\end{figure}

\newpage

\begin{figure}[htb]
 	\begin{center}
 		\includegraphics[width=0.98\textwidth]{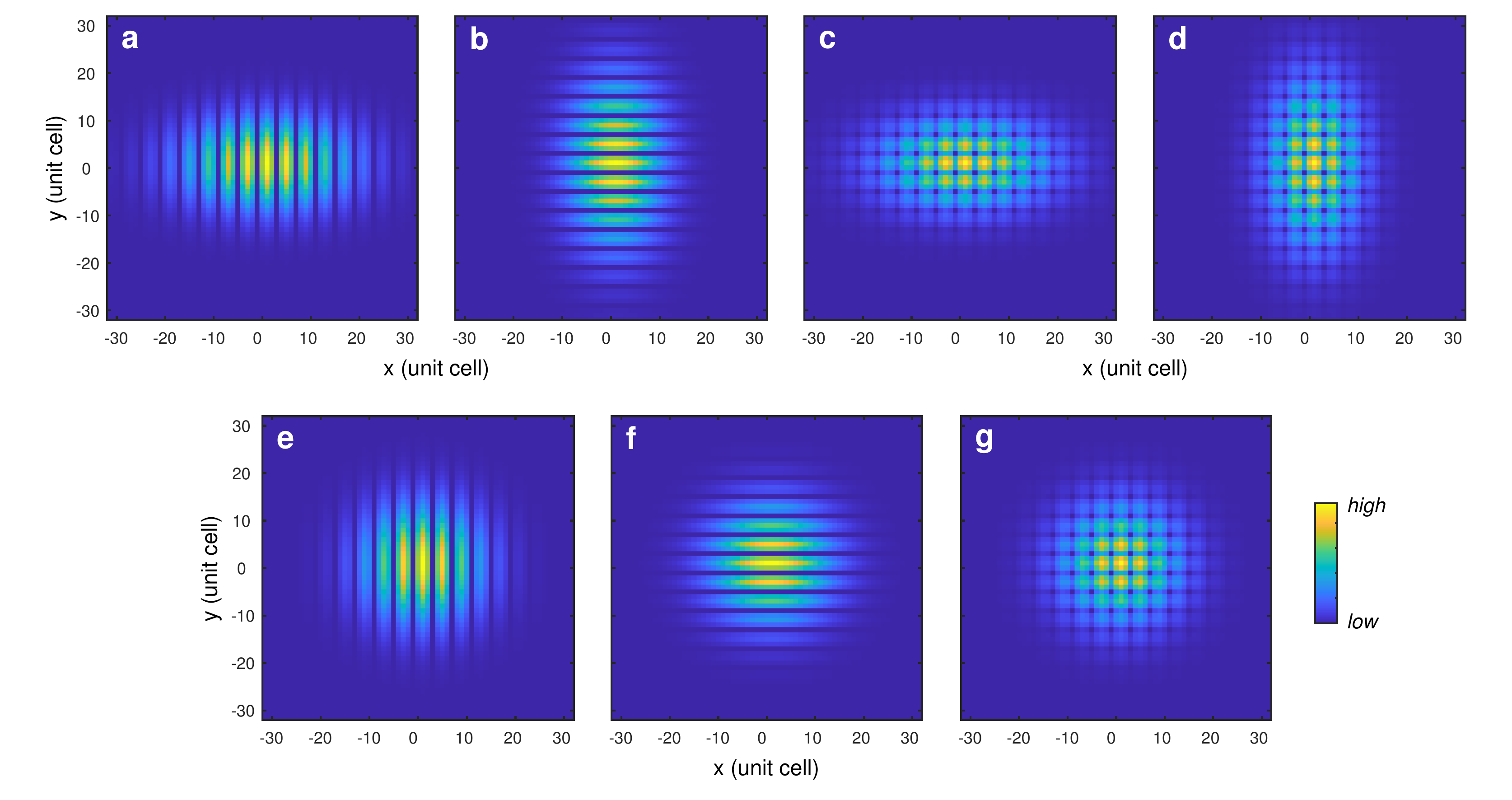}
 	\end{center}
 	\caption{\textbf{individual CDW domains.} \textbf{a},\textbf{b}, Electronic density modulation of an anisotropic stripe domain propagating into horizontal $x$ (\textbf{a}) and vertical $y$ direction (\textbf{b}) with four-unit-cell periodicity. \textbf{c}-\textbf{g}, Same for horizontally (\textbf{c}) and vertically anisotropic checkerboard domains (\textbf{d}), isotropic charge stripe domain propagating into horizontal $x$ (\textbf{e}) and vertical $y$ direction (\textbf{f}), and istoropic checkerboard domain (\textbf{g}). These domains are used as building blocks for modeling real-space electronic density map $\rho(x,y)$, illustrated in Supplementary Fig.~\ref{fig:S5}. } \label{fig:S4}
\end{figure}

\newpage

\begin{figure}[htb]
 	\begin{center}
 		\includegraphics[width=0.9\textwidth]{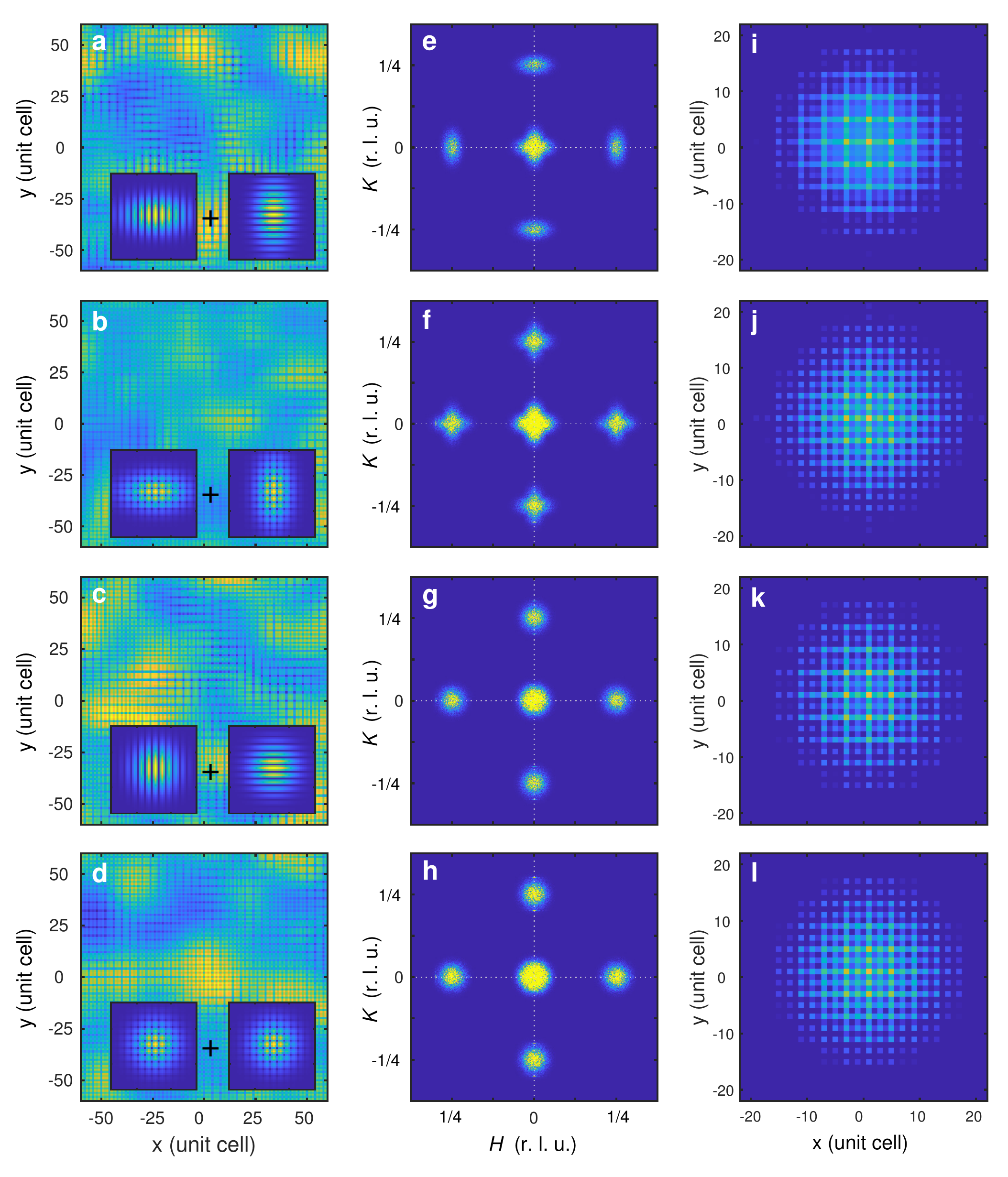}
 	\end{center}
 	\caption{\textbf{Modeling CDW modulation using simplified electron density maps.} \textbf{a}-\textbf{d}, Real-space electronic density modulation map $\rho(x,y)$ generated by equal population of individual charge stripe and checkerboard domains. The insets represent the microscopic structure of individual domains for each scenario. \textbf{e}-\textbf{h}, Structure factor $S(H,K)$ in reciprocal space, calculated by a direct Fourier transform of the corresponding $\rho(x,y)$ in (\textbf{a}-\textbf{d}), respectively. \textbf{i}-\textbf{l}, Density-density correlation map $C(x,y)$ directly computed by taking the auto-correlation of real-space map $\rho(x,y)$ in (\textbf{a}-\textbf{d}).} \label{fig:S5}
\end{figure}

\newpage

\begin{figure}[htb]
 	\begin{center}
 		\includegraphics[width=0.9\textwidth]{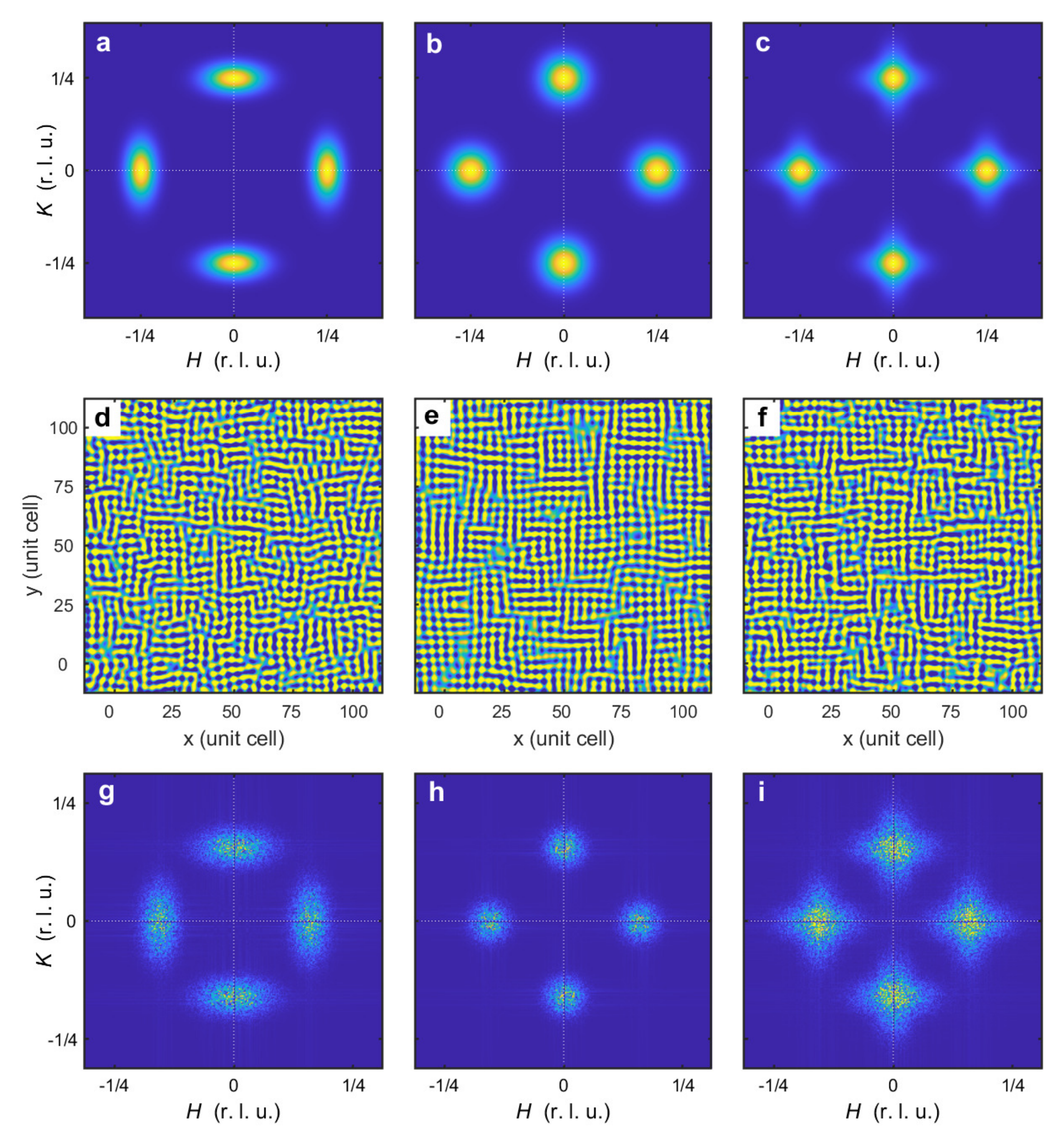}
 	\end{center}
 	\caption{\textbf{Modeling CDW modulation for more realistic electron density maps.} \textbf{a}-\textbf{c}, Structure factors $S(H,K)$ calculated for three different CDW peak patterns in reciprocal space. The parameters used in the calculation are summarized in Section II. \textbf{d}-\textbf{f}, Model electronic density map $\rho(x,y)$ reconstructed from the panel (\textbf{a}-\textbf{c}), respectively, under the assumption of random phase shifts. \textbf{g}-\textbf{i}, Fourier transformed intensity map of $\rho(x,y)$ in (\textbf{d}-\textbf{f}).} \label{fig:S6}
\end{figure}

 \end{document}